# Antiferromagnet–semiconductor van der Waals heterostructures: interlayer interplay of exciton with magnetic ordering

*Masaru Onga[1,2], Yusuke Sugita[2], Toshiya Ideue[1,2], Yuji Nakagawa[1,2], Ryuji Suzuki[1,2], Yukitoshi Motome[2], Yoshihiro Iwasa[1,2,3] ¹*

[1] Quantum-Phase Electronics Center (QPEC), The University of Tokyo, Tokyo 113-8656, Japan

[2] Department of Applied Physics, The University of Tokyo, Tokyo 113-8656, Japan

[3] RIKEN Center for Emergent Matter Science (CEMS), Wako 351-0198, Japan

ABSTRACT

Van der Waals (vdW) heterostructures have attracted great interest because of their rich material combinations.The discovery of two-dimensional magnets has provided a new platform for magnetic vdW heterointerfaces; however, research on magnetic vdW heterointerfaces has been limited to those with ferromagnetic surfaces. Here we report a magnetic vdW heterointerface using layered intralayer-antiferromagnetic $M$PSe$_3$ ($M$=Mn, Fe) and monolayer transition metal dichalcogenides (TMDs). We found an anomalous upshift of the excitonic peak in monolayer TMDs below the antiferromagnetic transition temperature in the $M$PSe$_3$, capturing a signature of the interlayer exciton-magnon coupling. This is a concept extended from single materials to heterointerfaces. Moreover, this coupling strongly depends on the in-plane magnetic structure and stacking direction, showing its sensitivity to their magnetic interfaces. Our finding offers an opportunity to investigate interactions between elementary excitations in different materials across interfaces and to search for new functions of magnetic vdW heterointerfaces.

TEXT

Apart from conventional heterostructures with III-V compound semiconductors or complex oxides, van der Waals (vdW) heterostructures attract much attention in multiple fields owing to their wide extensibility[1]. Because of the weak vdW force between their layers, we can freely combine any types of cleavable materials to make heterointerfaces regardless of their crystal structures. This provides a wide platform for making heterointerfaces with noteworthy physical properties: ultra-high mobility and quantum Hall effect[2]; vertical PN junctions[3]; and moiré physics related to Mott insulating states[4], superconductivity[5], and excitonic states[6–9]. Research on magnetic vdW heterointerfaces has greatly developed since the discovery of two-dimensional (2D) ferromagnets such as $CrGeTe_3$ and $CrI_3$[10,11]. In particular, an optical study of a monolayer $WSe_2$/$CrI_3$ heterostructure has clarified the interfacial magnetic coupling between a 2D nonmagnetic semiconductor and a ferromagnetic surface of the 2D magnetic insulator[12].

However, there has been no study focusing on antiferromagnetic properties (especially intralayer antiferromagnetic ordering) in magnetic vdW heterostructures, although antiferromagnets have recently been found effective for realizing spintronic functions[13]. Because antiferromagnets have a variety of spin orderings with distinct magnetic symmetry groups, we can anticipate unique magnetic properties of the heterointerfaces and control their functionalities by choosing appropriate magnets. The antiferromagnetic state is intrinsically free from stray fields, implying spintronic properties of the antiferromagnets are robust against external perturbation. More importantly, the energy gaps of the spin excitations in antiferromagnets are generally two or three orders of magnitude larger than those in ferromagnets. This can enhance the coupling between light and magnetic excitations in antiferromagnets from the viewpoint of opto-spintronics[14]. Constructing heterointerfaces with materials of high optical quality makes it possible to use the antiferromagnetic properties efficiently with other optical systems.

Here we report the optical properties of the vdW heterointerfaces of a layered intralayer-antiferromagnetic insulator ($M$PSe$_3$; $M$=Mn or Fe) and a nonmagnetic monolayer direct-gap semiconductor (MoSe$_2$) (Fig. 1a). The photoluminescence (PL) from the monolayer MoSe$_2$ is measured to probe the coupling between the antiferromagnet and the semiconductor via the vdW interface. We find an additional upshift of the excitonic peaks of MoSe$_2$ below the Néel temperature $T_N$, even though the antiferromagnetic ordering does not occur inside the MoSe$_2$ but in the neighbouring $M$PSe$_3$. The underlying origin of the shift cannot be explained by a homogenous exchange field like the Zeeman splitting in the ferromagnetic vdW interface[4,15]. Rather, our experimental and theoretical results suggest that the coupling is due to interlayer exciton–magnon coupling as shown in Fig. 1b: excitons in a semiconductor layer couple with magnetic excitations in the neighbouring antiferromagnet layer. The interlayer coupling can be turned on and off by intentionally selecting the magnetic ordering and stacking angle, which is enabled by the vdW nature of these heterointerfaces in marked contrast to single bulk magnets.

Monolayer transition metal dichalcogenides (TMDs) including $MoSe_2$ have been intensively studied and have excellent electrical/optical properties as direct-gap semiconductors with large exciton binding energy and Zeeman-type spin-orbit coupling[16,17]. In particular, excitonic peaks from monolayer $MoSe_2$ are separated enough to resolve the multiple excitonic states[18], and thus $MoSe_2$ is the most appropriate TMD for studying in detail the optical response at the interface for our purpose. For the antiferromagnet, we chose transition metal phosphorus trichalcogenides ($MPX_3$; $M$: transition metal; $X$: chalcogenides), which have been studied in bulk form with various intralayer antiferromagnetic orderings at $M$ such as the Néel-type in $MnPSe_3$ and the zigzag-type in $FePSe_3$ (the details are given in section §1 of the Supporting Information (SI))[19]. $MPSe_3$ is composed of magnetic ions ($M$) forming a honeycomb lattice within each layer and a $P_2Se_6$ ligand at the centre of the honeycomb (Fig. 1a). Each $M$ ion is octahedrally coordinated with six Se atoms from the ligands. Atomically thin $FePS_3$ was recently reported to show an antiferromagnetic phase transition even in the monolayer[20,21], suggesting that this family has magnetic ordering even in the thin-flake form. Note that $MPX_3$ possesses intralayer antiferromagnetic ordering in contrast to the interlayer antiferromagnetic ordering in multilayer $CrI_3$[11], meaning that our study is distinct from that on heterointerfaces with multilayer $CrI_3$[12] (which has the ferromagnetic intralayer ordering at each layer).

We fabricated vdW interfaces of monolayer $MoSe_2$ and multilayer $MPSe_3$ with a thickness of tens of nanometers. The interfaces were fabricated via a mechanical exfoliation and an all-dry transfer method[22] inside a glovebox with an inert atmosphere to prevent sample degradation (Fig. 2a). We controlled the stacking angles according to the relation between the edge of the cleaved flakes and the crystal direction[23]. The zigzag-edge of $MoSe_2$ was aligned parallel to the zigzag-edge of honeycomb $Mn^{2+}$ (sample A1) to realize nearly commensurate stacking between the $2 \times$ 2 superlattice of MoSe$_2$ (lattice constant $a_{MoSe2}$ = 0.328 nm) and the unit cell of $M$PSe$_3$ ($a_{MnPSe3}$ = 0.639 nm) with a lattice mismatch of 2.6% (see the details in SI section §2).

Figure 2b is the Raman spectrum of a typical heterostructure, monolayer MoSe$_2$/thin MnPSe$_3$. Clear MnPSe$_3$ peaks appear at 149 cm$^{-1}$, 175 cm$^{-1}$, and 222 cm$^{-1}$, which is consistent with a previous study[24]. There is also a relatively weak peak of monolayer MoSe$_2$ at 244 cm$^{-1}$. The Raman intensity of MnPSe$_3$ is known to be enhanced below $T_N$[24], and can thus be an optical probe of an antiferromagnetic transition in the exfoliated thin flakes. Figure 2c (and Fig. S2 in SI section §3) shows the temperature dependence of the main peak of thin MnPSe$_3$ at 222 cm$^{-1}$, indicating that the Raman intensity is also enhanced even in exfoliated thin flakes. The upturn starts around $T_N$=74 K as determined by a previous study on neutron scattering[19], meaning that the magnetic orderings in our samples are not affected by thinning at least down to ~20 nm.

Figure 2d shows the PL spectra of monolayer MoSe$_2$ on MnPSe$_3$ and on SiO$_2$ at 6 K. Each spectrum has a neutral exciton (X$^0$) peak around 1.66 eV and a trion (X$^T$) peak around 1.62 eV. The MoSe$_2$ peaks on MnPSe$_3$ are slightly broader than those on SiO$_2$ owing to photocarrier relaxation from MoSe$_2$ to MnPSe$_3$, as observed in a ferromagnetic vdW interface with spin relaxation[12]. The differences in peak positions between the two samples are due not only to strain induced by the substrates, but also to the magnetic ordering of the bottom material as we discuss below. All the typical spectra in this study and detailed discussion of the peaks are in section §4 and Fig. S3 of the SI.

We performed PL measurements at various temperatures to clarify the effect of the magnetic ordering on the spectra. These spectra are shown in Fig. 3a for MoSe$_2$/MnPSe$_3$ (sample A1) and MoSe$_2$/SiO$_2$ (sample S1). The MoSe$_2$/SiO$_2$ spectra in Fig. 3a are uniformly shifted by +7 meV, a temperature-independent difference due to the constant strain from the substrates, because we aim to uncover only the effect of the magnetic ordering of MnPSe$_3$. Hereafter, we focus on the X$^0$ peaks because we can clearly observe them at any temperature. The X$^0$ peak positions are indicated with triangles in Fig. 3a and plotted in Fig. 3b as a function of temperature. The temperature dependence of MoSe$_2$/SiO$_2$ follows the well-known behaviour of semiconductor band gaps (see SI section §5)[18]. While both peaks are upshifted together on cooling above $T_N$ for MnPSe$_3$, we found they are separated on further cooling below $T_N$. Figure 3c is a plot of the difference between the two peak positions against temperature. The marked feature in crossing $T_N$ strongly suggests that the magnetic ordering inside MnPSe$_3$ affects the exciton of the attached monolayer MoSe$_2$. Similar results are also observed in other samples with MoSe$_2$ (samples A2 and A3 in SI section §6), WSe$_2$, and MoS$_2$ (shown in SI section §7); therefore the observed shift is common in semiconducting group-VI TMDs on MnPSe$_3$. These results for antiferromagnetic heterointerfaces, however, have never been observed or predicted to our knowledge despite much work on heterointerfaces with ferromagnetic surfaces[12,15,25].

Exciton peaks in bulk semiconducting antiferromagnets, on the other hand, are known to be modulated below $T_N$, which is attributed to exciton-magnon coupling (EMC)[26]. This means that an exciton in an antiferromagnet itself can directly interact with a magnon via EMC. The energy of an absorbed/emitted photon corresponds to the addition/subtraction of exciton and magnon energies, meaning that the EMC can cause optical peak modulation equal to the magnon energy[26]. Absorption spectra of bulk $MPX_3$ crystals reveal an EMC effect in d-d transitions[27] and additional upshifts in p-d transitions below $T_N$[28]. Similar phenomena have been also reported on other antiferromagnetic insulators[29-31]. We can therefore interpret the present results as exciton–magnon coupling across the interface in analogy: the exciton in the semiconductor layer couples with the magnon in the antiferromagnetic layer through interlayer interaction (Fig 1b and inset of Fig. 3c).

Hereafter, we call this interlayer exciton–magnon coupling (interlayer EMC) to distinguish it from conventional EMC at the same atomic sites. Note that the observed energy shift in Fig. 3c is roughly comparable to the magnon energy scale in bulk $MnPSe_3$ (several meV) measured via inelastic electron tunnelling spectroscopy in a previous study[32].

Meanwhile, we considered three possible origins other than the interlayer EMC: the band modulation of monolayer $MoSe_2$ by the antiferromagnetic order in $MnPSe_3$, the additional strain due to magnetostriction, and the bound magnetic polaron (shown in detail in SI sections §8–11). In particular, we conducted density functional theory (DFT) calculations for an exactly commensurate TMD/$MnPSe_3$ system (corresponding to the ideal configuration of sample A1) to find out whether the band structure of monolayer TMDs can be directly modulated through interlayer coupling with antiferromagnets. We concluded, however, that none of these three possibilities could explain our experimental results, and thus attributed them to the interlayer EMC in analogy to the mechanism in bulk systems.

Next, we demonstrate that the interlayer EMC can be turned on in a particular vdW configuration. We conducted the same measurements on $FePSe_3$ as on $MnPSe_3$ to reveal the effect of magnetic orderings. $MnPSe_3$ shows a Néel-type antiferromagnetic ordering (Fig. 4a) below $T_N = 74$ K, whereas a zigzag-type ordering occurs in $FePSe_3$ below $T_N = 112$ K (Fig. 4b)[19], causing distinct coupling with the top $MoSe_2$ layer from $MnPSe_3$. Figure 4c shows the $X^0$ peaks of $MoSe_2$/$FePSe_3$ (sample F1) against temperature along with those of $MoSe_2$/$MnPSe_3$ and $MoSe_2$/$SiO_2$. The temperature dependence of $MoSe_2$/$FePSe_3$ is more similar to that of $MoSe_2$/$SiO_2$ than that of $MoSe_2$/$MnPSe_3$, indicating that the zigzag-type antiferromagnetic ordering of $Fe^{2+}$ affects the exciton of the neighbouring TMDs much more weakly than the Néel-type ordering of $Mn^{2+}$.

Additionally, we inspected the effect of the stacking angle between MoSe$_2$ and MnPSe$_3$ layers, which can give more microscopic information about the interlayer EMC. We fabricated another sample with a different stacking angle than that of sample A1: sample E1 with the zigzag-edge of MoSe$_2$ perpendicular to the zigzag-edge of Mn$^{2+}$ (⊥). In contrast to the parallel configuration in sample A1, the perpendicular one is not expected to have good lattice matching, resulting in a moiré pattern and an additional PL peak possibly from a moiré exciton (see SI section §2). The temperature dependence of the sample E1 peaks in Fig 4d shows that the parallel configuration (∥) is essential for the additional peak shifts below $T_N$. This dependence on stacking direction and the above comparison with FePSe$_3$ (summarized in Fig. 4e including MoSe$_2$/FePSe$_3$ (⊥)) strongly suggest that interlayer EMC is very sensitive to the types of magnetic ions (Mn$^{2+}$ or Fe$^{2+}$) and/or interfacial symmetry (magnetic ordering patterns or the crystal direction). This indicates that it is possible to turn on the EMC via vdW engineering, which is a unique nature of interlayer exciton-magnon coupling and was impossible to investigate in single bulk AFMs. We note that such sensitivity has not been expected for ferromagnetic vdW interfaces, in which the exchange interaction is basically uniform irrespective of stacking angle.

The detailed theoretical understanding of this interlayer EMC remains to be established (see SI §12). The original theory on exciton-magnon coupling in bulk (formula (2) of SI §12) is based on the electric dipole moment generated by the combination of an orbital excitation (exciton) at one sublattice and a spin excitation (magnon) at the other sublattice *via* exchange interaction between the sites[26]. In the interlayer EMC which we propose here, we suppose that interlayer exchange interaction connects an exciton in the TMDs layer and a magnon in the antiferromagnetic layer. This conceptual expansion can make it complicated and/or interesting to construct the microscopic theory. From the other points of view, tight-binding approach including interlayer hopping can give us suggestions on this system similarly to the discussion on the interlayer exciton (interlayer coupling between charged particles). Experimentally, it would be helpful to excite the magnons directly by microwave or THz light in order to prove the EMC directly. Detailed properties can be revealed by investigating stacking-angle dependence and other material combinations.

From a different perspective, interlayer coupling between elementary excitations is intriguing both for basic science and applications. It would be fascinating to understand how the valley-excitons (exciton with valley degrees of freedom) microscopically interact with magnetic excitations. Efficient magnon-photon (microwave-light) transduction for opto-spintronics/quantum-electronics could also be expected by regarding monolayer TMDs with the interlayer EMC as a repeater of magneto-optical coupling of the magnetic insulators.

In conclusion, we fabricated antiferromagnet–semiconductor van der Waals heterointerfaces and found an anomalous upshift of excitonic peaks, which is presumably induced by interlayer exciton–magnon coupling. These features suggest that the interaction between elementary excitations such as excitons and magnons, which has been studied for single substances, can be conceptually expanded to artificial heterointerfaces and modulated through vdW engineering. These results open a realm of magnetic vdW interfaces with versatile functions for opto-spintronics and quantum electronics and provide detailed insight into correlated excitations in condensed matter physics.

FIGURES

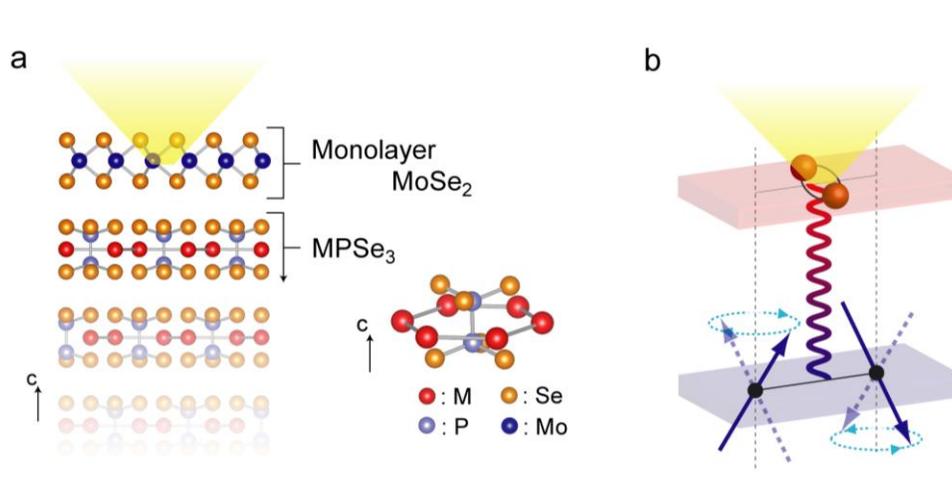

**Figure 1**. **Antiferromagnet–semiconductor van der Waals herointerfaces. a**, Crystal structure of our system. The heterostructure is made of two layered materials: monolayer $MoSe_2$ and tens-of-nanometers-thick $MPSe_3$. The $MPSe_3$ is composed of a $M^{2+}$ plane with honeycomb structure and a $[P_2Se_6]^{4-}$ dimer. **b**, Coupling between an exciton in one layer and a magnon in the other layer via interlayer interaction. Our experimental data suggest that both excitations interact with each other over the van der Waals gap.

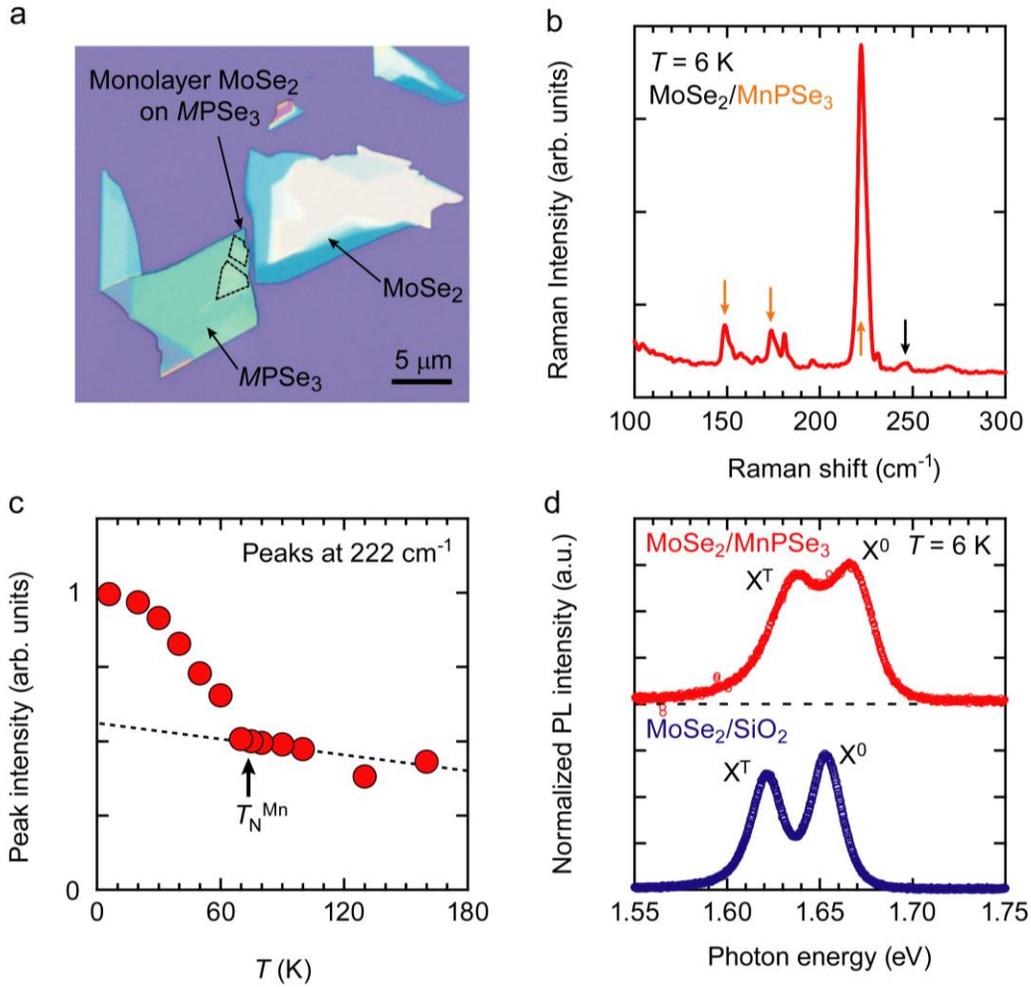

**Figure 2. Characterization of a monolayer MoSe$_2$/MnPSe$_3$ heterointerface. a**, Optical image of a heterostructure on a SiO$_2$/Si substrate. Monolayer MoSe$_2$ on MnPSe$_3$ is highlighted by the black dotted lines **b**, Typical Raman spectrum of the heterostructure, which includes strong peaks (orange arrows) from MnPSe$_3$ and a weak peak (a black arrow) from monolayer MoSe$_2$. **c**, Temperature dependence of the peak intensity at 222 cm$^{-1}$ in Fig. 2b. The sudden increase of the intensity indicates the tens-of-nanometers-thick MnPSe$_3$ has an antiferromagnetic transition at almost the same critical temperature as the bulk. The dashed line is a guide for the eyes. **d**, Photoluminescence (PL) spectra of MoSe$_2$/MnPSe$_3$ (red) and MoSe$_2$/SiO$_2$ (blue) at 6 K. Both spectra include neutral exciton (X$^0$) and trion (X$^T$) peaks.

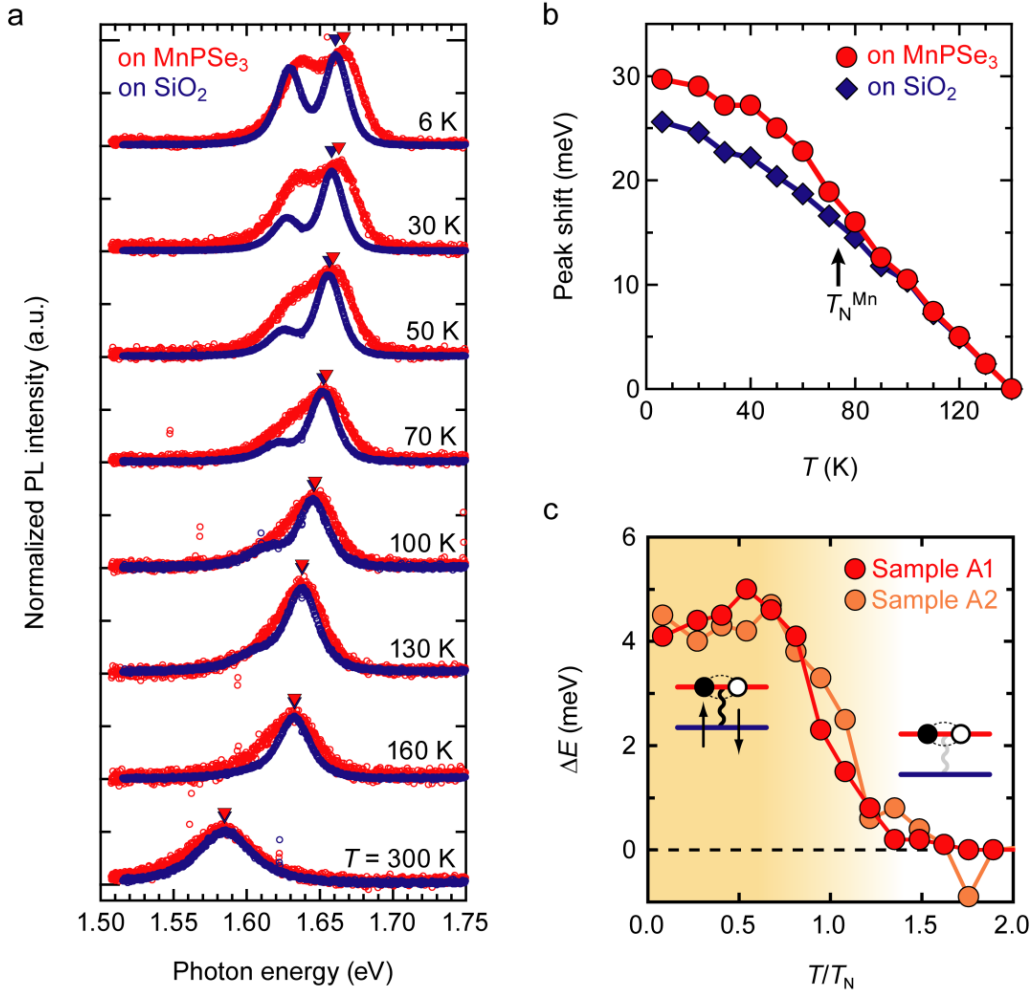

**Figure 3. Temperature dependence of PL peaks. a**, PL spectra at various temperatures. All MoSe$_2$/SiO$_2$ spectra are uniformly shifted by +7 meV for comparison. The triangles indicate the peak positions of X$^0$. **b**, Temperature dependence of the peak positions of MoSe$_2$/MnPSe$_3$ (red circles; sample A1) and MoSe$_2$/SiO$_2$ (blue squares; sample S1) in Fig. 3a. The values at 140 K are set as the origins of the shifts. **c**, Peak energy shifts of two MoSe$_2$/MnPSe$_3$ samples due to antiferromagnetic ordering, where Δ$E$ is the difference between the peak shifts of $M$PSe$_3$ and that of SiO$_2$ at each temperature. Temperature on the horizontal axis is normalized by $T_N$ for $M$PSe$_3$. The left and right insets show interlayer couplings at the interface between excitons in a semiconductor (red) and an antiferromagnet (blue) below and above $T_N$, respectively.

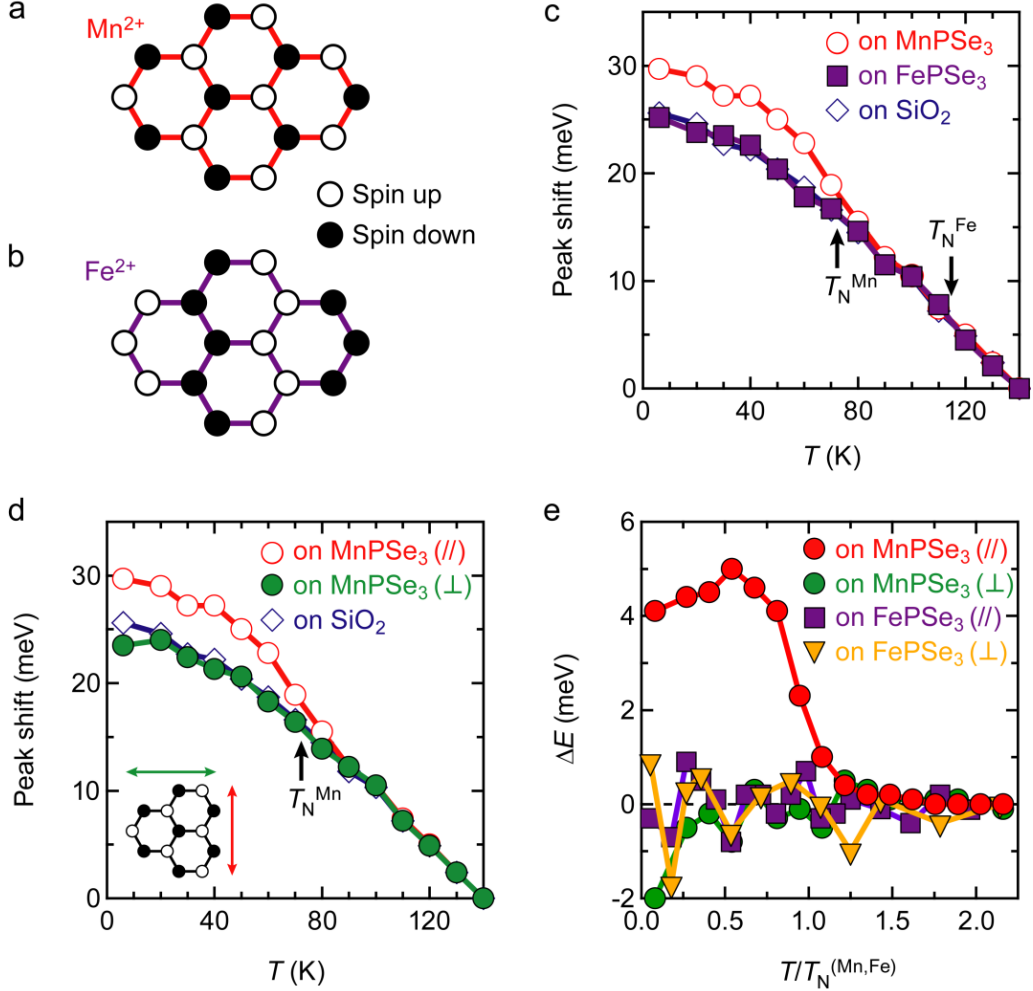

**Figure 4. Magnetic-order and stacking-pattern dependence. a, b,** Antiferromagnetic ordering of $M^{2+}$ in $M$PSe$_3$. Each spin lies within the basal plane in bulk MnPSe$_3$ and is parallel to the c-axis in bulk FePSe$_3$ (see SI section §1). **c,** Temperature dependence of the peak shift in different $M^{2+}$ structures. The red open circles and purple squares show the peak shifts of the heterostructures consisting of Néel-type Mn$^{2+}$ (sample A1) and zigzag-type Fe$^{2+}$ (sample F1), respectively, and the blue open diamonds indicate the shifts for MoSe$_2$/SiO$_2$ (sample S1). **d,** Temperature dependence of peak shifts in different stacking patterns. The red open circles, green circles, and blue open diamond represent the PL peaks of MoSe$_2$ on MnPSe$_3$ with parallel configuration (∥; sample A1), on MnPSe$_3$ with perpendicular configuration (⊥; sample E1), and on SiO$_2$ (sample S1),

respectively. **e,** Summary of the experimental data. The red circles, green circles, and purple squares indicate $\Delta E$ for $MoSe_2/MnPSe_3$ (∥; sample A1), $MoSe_2/MnPSe_3$ (⊥; sample E1), $MoSe_2/FePSe_3$ (∥; sample F1), and $MoSe_2/FePSe_3$ (⊥; sample G1) respectively.

METHODS

**Sample preparation:** Single crystals of $MoSe_2$, $MnPSe_3$, and $FePSe_3$ were grown via a chemical vapour transport technique[33,34]. Monolayer $MoSe_2$ flakes were exfoliated directly onto polydimethylsiloxane (PDMS), and $MPSe_3$ flakes were exfoliated onto $SiO_2$(300 nm)/Si substrates in a glovebox filled with $N_2$. We then transferred monolayer $MoSe_2$ onto a $MPSe_3$ flake on the substrate (using an all-dry transfer method[22]) in the glovebox, put the sample into the chamber, and flew $Ar/H_2$ (97%/3%) gas at 523 K for 2 hours to fit the flakes to each other. The thickness was determined via optical contrasts and atomic force microscopy images.

**Photoluminescence measurement:** We performed the optical measurements in a He-flow cryostat (Oxford Instruments) under a high vacuum (~$5\times10^{-6}$ torr) at low temperature. Raman spectra were recorded on a JASCO spectrometer equipped with a semiconductor laser (532 nm). In photoluminescence spectroscopy, we used a spectrometer with a liquid-nitrogen-cooled CCD (Horiba) and a He–Ne laser (632.8 nm; 5 mW; Melles Griot). The power was decreased to 500 µW, and the beam was focused onto the monolayer $MoS_2$ flakes using a 50x objective (Olympus).

**DFT calculation:** In DFT calculations, we used the OpenMX code[35], which is based on a linear combination of pseudoatomic orbital formalism[36,37]. We used the Perdew-Burke-Ernzerhof generalized gradient approximation (GGA) functional in density functional theory[38], the DFT-D2 method to take van der Waals force into account[39], and a 16×16×1 k-point mesh and vacuum space greater than 20 Å between bilayer systems of TMD and $MnPSe_3$ for the calculation of the self-consistent electron density and the structure relaxation. We optimised the atomic positions in the unit cell with the convergence criterion $10^{-2}$ eV/Å for the inter-atomic forces starting from the stacking structure concluded as the most stable one in the literature[40]; we confirmed that all the results stably converged to similar lattice structures. See the Supporting Information for further details of the DFT calculations.

**Supporting Information**. The following files are available free of charge.

Supporting_information.pdf (including supplementary discussions related to this letter)

AUTHOR INFORMATION

**Corresponding Author**

*E-mail: iwasa@ap.t.u-tokyo.ac.jp

**Notes**

The authors declare no competing financial interest.

ACKNOWLEDGMENT

We thank K. Usami and A. Fujimori for helpful discussions. Some calculations were carried out at the Supercomputer Center of the Institute for Solid State Physics at The University of Tokyo. M.O., Y.S., and Y.N. were supported by the Japan Society for the Promotion of Science (JSPS) through the Research Fellowship for Young Scientists. T.I. was supported by JSPS KAKENHI grant numbers JP19K21843, JP19H01819, and JST PRESTO project (JPMJPR19L1). This research was supported by a Grant-in-Aid for Scientific Research (S) (No. 19H05602) and the A3

# Supporting Information:

## Antiferromagnet–semiconductor van der Waals heterostructures: interlayer interplay of exciton with magnetic ordering

Masaru Onga[1,2], Yusuke Sugita[2], Toshiya Ideue[1,2], Yuji Nakagawa[1,2], Ryuji Suzuki[1,2], Yukitoshi Motome[2], Yoshihiro Iwasa[1,2,3] *

[1] Quantum-Phase Electronics Center (QPEC), The University of Tokyo, Tokyo 113-8656, Japan
[2] Department of Applied Physics, The University of Tokyo, Tokyo 113-8656, Japan
[3] RIKEN Center for Emergent Matter Science (CEMS), Wako 351-0198, Japan
*Correspondence to: iwasa@ap.t.u-tokyo.ac.jp

§1. Magnetic structures of $M\mathrm{P}X_3$

$M\mathrm{P}X_3$ family ($M$: transition metals, $X$: chalcogenides) shows various types of magnetic ordering[1]. The spin orderings occur at honeycomb $M^{2+}$ within the layer: for instance, $MnPS_3$ has the Néel-type spin ordering parallel to c-axis at $Mn^{2+}$ ions, while $FePS_3$ and $FePSe_3$ have the zigzag-type orderings also along c-axis at $Fe^{2+}$ ions.

The magnetic moment of the Néel-type $MnPSe_3$ is supposed to lie within the basal plain (or slightly canted between the plain and c-axis) according to the neutron diffraction study on polycrystalline powder[2]. In our calculation shown later in §9, however, the monolayer $MnPSe_3$ beneath the monolayer TMDs has an easy axis along c-axis similarly to the previous first-principal calculation on a bilayer $MnPSe_3$[3]. We have no conclusive picture of the actual spin ordering at the surface of $MnPSe_3$ due to the above controversy and the experimental difficulty to detect it.

We can guess that the heterostructure made of $MnPS_3$ (with Neel-type $Mn^{2+}$) would show the similar effects to that of $MnPSe_3$. However, the situation could be dramatically distinct because of the different orientations of sublattice spins in bulk $MnPS_3$ (along c-axis) and bulk $MnPSe_3$ (almost lied in ab-plane) as we noted above. Since the magnetoresponse of the TMDs is quite anisotropic owing to the Zeeman-type spin-splitting at the K-points, the excitons at the interface with antiferromagnets can also be sensitive to the spin-orientation of the magnets as well as the types of spin-ordering. It would be very informative to explore the detailed experiments with other $M$P$X_3$ in order to elucidate the microscopic coupling at the interface. The direction of the sublattice magnetic moments can also explain the $M$-dependency of the exciton-magnon interaction discussed in Fig.4 of the main text.

§2. Commensurability depending on stacking angles

The $2 \times 2$ superlattice of MoSe$_2$ (lattice constant $a_{\text{MoSe2}} = 0.328$ nm) and the unit-cell of $M$PSe$_3$ (exactly speaking $M_2$P$_2$Se$_6$, $a_{\text{Mn2P2Se6}} = 0.639$ nm and $a_{\text{Fe2P2Se6}} = 0.632$ nm) present nearly commensurate lattice matching with the parallel configuration: their lattice mismatches are 2.6% in MnPSe$_3$ and 3.6% in FePSe$_3$ as calculated from the lattice constants of bulk. Figure S1a shows the image of MoSe$_2$/MnPSe$_3$ in the parallel case. In contrast, we cannot expect the good lattice matching with the perpendicular configuration as shown in Fig. S2b, visually presenting moiré patterns in this case within this scale. In the parallel case, we cannot see moiré patterns in the scale of Fig. S1a (~ 20 unit cells of MoSe$_2$) due to the relatively small lattice mismatch of 2.6%, meaning that more than 80 unit cells are needed to see such a periodic pattern.

§3. Detailed discussions on Raman spectra

We show Raman spectra around 222cm$^{-1}$ at various temperatures in Fig. S2 as raw data of Fig. 2c in the main text. The abrupt changes around $T_N$=74 K of the bulk MnPSe$_3$ support that the antiferromagnetic transition of the exfoliated $MnPSe_3$ occurs at almost same temperature as the bulk.[2,4]. Similar behavior is also observed in the peaks around 149 and 175 cm$^{-1}$.

Generally speaking, two-magnon scattering can cause a broad Raman peak at the two-magnon frequency at Brillouin zone boundary (~9 meV × 2 = 145 cm$^{-1}$ in $MnPSe_3$ [3]). A recent study on Raman spectroscopy of bulk/exfoliated $MnPS_3$ reports a small two-magnon peak above 2.71 eV-excitation in a bulk sample[5]. However, no two-magnon peak has been observed within our measurements with 2.33 eV-excitation. Considering the previous work mentioned above[5], it could be due to the small excitation energy and/or weaker intensity of the two-magnon peak in exfoliated flakes than a prominent phonon peak around 149 cm$^{-1}$.

§4. PL spectra from $MoSe_2$/$M$PSe$_3$: Possible moiré excitons and strain from substrates

Figure S3 display four PL spectra of sample F1 ($MoSe_2$/$FePSe_3$, ∥), A1 ($MoSe_2$/$FePSe_3$, ∥), S1 ($MoSe_2$/$SiO_2$), E1 ($MoSe_2$/$MnPSe_3$, ⊥), and G1 ($MoSe_2$/$FePSe_3$, ⊥) at 6 K. We also plot the fitting results by multi-peaks Voigt function: the spectra of sample F1, A1 and S1 can be decomposed well into two peaks, while that of sample E1 and G1 includes more peaks. Here we discuss the origins and positions of the peaks in detail.

Following the previous study on $MoSe_2$/$SiO_2$[6], the peaks at the highest energy in each spectrum are attributed to the neutral excitons ($X^0$), while the peaks located 30 meV lower than $X^0$ are assigned as the charged excitons (trions, $X^T$). Only in the spectrum of the perpendicular configuration of $MoSe_2$/(Mn, Fe)$PSe_3$ (sample E1, E2 and G1), we found an unknown peak ($X^M$) located at 10 ~ 15 meV lower than $X^0$. This middle peak $X^M$ is assigned as neither trions nor neutral biexcitons, because the binding energies from the $X^0$ state are 30 meV both for positively and negatively charged trions regardless of the doping level[6] and 20 meV for neutral biexcitons[7]. This suggests that the $X^M$ is likely attributed to moiré excitons affected by the moiré potential at the interface (shown in Fig. S2b) as discovered in TMD-TMD heterostructures[8-11].

Regarding sample G1 (MoSe$_2$/FePSe$_3$, perpendicular) in Fig. S3, the spectrum appears complicated and includes multiple peaks, some of which are absent in sample F1 (MoSe$_2$/FePSe$_3$, parallel). For peak extraction, we assumed here that the spectrum of G1 should be composed of $X^0$ and $X^T$ which exist all other spectra (with the constant trion binding energy (35 meV)), and then found to be decomposed into four peaks ($X^0$, $X^T$, and other two peaks) by fitting. One of them is located at $X^M$ of sample E1, so we also named it $X^M$ of G1 after that of E1. The last and lowest peak is unknown, but we assigned it as localized exciton ($X^L$). $X^L$ has been frequently discussed in MoS$_2$ (or WSe$_2$) as a defect-mediated bound state.

The position of each peak can be explained by the lattice mismatch and the consequent innate strain except the antiferromagnetic effect which is mainly discussed in the main text. If the lattice relaxation occurs at the present vdW interface, $M$PSe$_3$ induces the compressive strain to the monolayer MoSe$_2$ in the parallel configuration, while the lattice mismatch on the SiO$_2$ substrate can be regarded as zero because the surface structure of SiO$_2$ is amorphous. Lattice reconstruction at the van der Waals interfaces has been directly reported in graphene/BN[12] and graphene/graphene[13] systems although such reconstruction at vdW interfaces had been assumed to be absent at the early stage of this field. Our fabrication procedure includes annealing at 250 °C (same condition as the report showing the reconstruction at the heterointerface[12]), thus we can expect such lattice relaxation producing strain from the substrates.

Since the compressive strain enlarges the bandgap of TMDs[14,15], this mechanism can explain the substrate dependence of the $X^0$ and $X^T$ positions in Fig. S3; the peak positions line up in order of the lattice mismatch, sample S1 < A1 < E1. According to previous studies[14,15], the compressive strain is deduced from the observed peaks as 0.5 ~ 1%. The value is less than the lattice mismatch probably because the lattice of the surface of the counterpart $M$PSe$_3$ can be expanded. Sample E1 shows the peaks of X$^0$ and X$^T$ at similar positions as sample S1 and thus implies negligibly small strain in MoSe$_2$, being consistent with the nearly incommensurable picture in the perpendicular configuration as shown in Fig. S1b. Note here that we assume the effect of dielectric environment can be negligible because the peak position of MoSe$_2$/MnPSe$_3$ ($\perp$) is not shifted as seen in Fig. S3 even though the dielectric constant of bottom MnPSe$_3$ is identical to MoSe$_2$/MnPSe$_3$ (//).

§5. Temperature dependence of the X$^0$ peaks in MoSe$_2$ on SiO$_2$

The bandgaps of MoSe$_2$ and other TMDs are known to follow the temperature dependence of standard semiconductors as following[16]:

$$E(T) = E_0 - S\hbar\omega \left[\coth\left(\frac{\hbar\omega}{2k_BT} - 1\right)\right] \quad (1)$$

where $E_0$ is the bandgap (exciton energy) at zero temperature, $S$ is dimensionless coupling constant, and $\hbar\omega$ is an average phonon energy. Figure S4 shows the temperature dependence of sample S1 (MoSe$_2$ on SiO$_2$) and the fitting curve by (1). We deduced $E_0$ = 1.6514 eV, $S$ = 1.86, and $\hbar\omega$ = 14 meV, which are similar to the previous report (1.657 eV, 1.96, 15 meV, respectively)[6]. In the main text, we discuss the anomalous peak shifts in addition to this basic temperature dependence.

§6. Temperature dependence in other MoSe$_2$/$M$PSe$_3$ samples

To check reproducibility of the additional shifts below $T_N$, we fabricated and measured several samples in addition to sample A1, E1, F1, G1 and S1 in the main text. Figure S5 indicating the data from sample A2, A3, E2, S2 totally agrees to the data in Fig. 4, supporting our results are truly intrinsic in the system.

§7. Experimental data of WSe$_2$/$M$PSe$_3$ and MoS$_2$/$M$PSe$_3$ heterointerfaces

We also conducted the experiments using TMDs other than MoSe$_2$ in order to check the universality of our results among group-VI TMDs. The data shown in Fig. S6 indicate that the phenomena observed in MoSe$_2$/$M$PSe$_3$ occur also in WSe$_2$/$M$PSe$_3$ and MoS$_2$/MnPSe$_3$ and their features are similar quantitatively and qualitatively among TMDs. It is not easy to discuss their differences among TMDs in detail due to the differences in their lattice constant/band structure/spin-orbit interaction at present.

We note that the extracted peaks from WSe$_2$ and MoS$_2$ can include the luminescence from charged excitons and/or localized excitons due to their broadness and complexity of PL peaks while PL peaks from MoSe$_2$ are sharp enough to separate them clearly into neutral and charged excitons. Especially, further researches on the exciton-magnon interaction in charged excitons could give us a tool to tune the interaction electrically.

§8. Electronic band structure of the heterostructure by DFT calculations

We performed band structure calculations based on the density functional theory (DFT) for the bilayer system composed of monolayer TMD/monolayer MnPSe$_3$. The unit cell used in the calculations is shown in Figs. S7a and b, where we assumed the two Mn atoms are located beneath the Mo and Se atoms, respectively, to model the nearly commensurate stacking in the case of the parallel configuration discussed in the main text. We adopted MoS$_2$ as a monolayer TMD because it has a small lattice mismatch ($\sim$ 1.0%) with MnPSe$_3$[17]; we fixed the in-plane lattice constant of the unit cell to 6.390 Å and optimized the atomic positions in the unit cell. In the calculations, we employed the GGA+$U$ method taking $U = 5$ eV for the $d$ orbitals of Mn, assuming the Néel-type antiferromagnetic order with the magnetic moments along the out-of-plane directions in the honeycomb lattice formed by Mn sites.

The Brillouin zone (BZ) for this system is shown in Fig. S7c. The BZ of the superlattice (black rectangle in Fig. S7c) is one fourth of the original one of MoS$_2$ (orange rectangle), and hence, the high-symmetric points are defined in the folded positions (e.g., the K and M points indicated by the green letters). Note that the new K point originates from the "−K point" of the original BZ of MoS$_2$, and the new Γ-K-M line does not correspond to "the original Γ-K-M line" [e.g., the new Γ-K line does not include the "Q (or Λ) points" of the original MoS$_2$].

Figure S7d displays the band structure of the MoS$_2$/MnPSe$_3$ bilayer. For each electronic band, we indicate the contributions from the $d$ orbitals of Mo and Mn separately by the colored circles. Hereafter, all the origins of the energy are set at the top of the valence bands of MoS$_2$. The bottoms/tops of the conduction/valence bands from MoS$_2$ are isolated from the ones of MnPSe$_3$. We note that our results agree well with the previous results on the same system[17].

§9. Electronic band structure for different magnetic structures

We conducted three kinds of calculations to clarify how the bandgap of MoS$_2$ is modified by changing the magnetism on the neighboring MnPSe$_3$ layer. Specifically, we controlled the direction and amplitude of the magnetic moments, and type of the magnetic ordering of Mn$^{2+}$.

First, Fig. S8 shows the band structures for different directions of the magnetic moments of the Néel order, **m**: parallel to the c axis (**m**//c) and vertical to the c axis (**m**⊥c) in Figs. S8a and b, respectively. From the comparison of the total energies, the case with **m**//c is slightly more stable than that with **m**⊥c, as in the ground state of a bilayer MnPSe$_3$[2]. We find that the difference of the bandgaps of MoS$_2$ between the two cases is very small, less than 0.5 meV, as shown in Fig. S8c. This is one order of magnitude smaller than the PL spectral shift in the main text. We note that in both cases, the conduction band edges of MoS$_2$ exhibit the Zeeman-type spin polarization (almost aligned along the c-axis).

Next, in Fig. S9, we show the band structures with different amplitudes of the magnetic moments, $m=|\mathbf{m}|$, for the case with **m**//c. The results were obtained by the GGA+$U$ calculations with the constraints on the magnetic moments[18]. The bandgaps of MoS$_2$ are hardly modified by the change in the magnetic moments from 4.60 $\mu_B$ to 5.73 $\mu_B$, as shown in Fig. S9b.

Finally, we compare the band structures for the Néel and zigzag-type antiferromagnetic order in Fig. S10. The calculations are done by assuming the double supercell to incorporate the zigzag-type order (Figs. S10a and b). Note that a similar zigzag order is realized in FePSe$_3$. Here, we adopted not GGA+$U$ but GGA calculations because of the high computational cost for 40 atoms in the supercell. As shown in Fig. S10e, we find no significant difference in the bandgaps of MoS$_2$ for the different types of antiferromagnetic orders. Note that the additional bands, whose band bottom is lower than 1.570 eV, predominantly originate from Mn $d$ orbitals.

From these results, we conclude that the antiferromagnetic order in MnPSe$_3$ does not affect the bandgap of MoS$_2$ significantly. The change of the bandgap for different antiferromagnetic structures is too small to explain the upshift of the PL spectra in the main text. This is in stark contrast to the substantial change in the bandgap for the ferromagnetic substrate suggested theoretically[19]. The band alignment would be changed before and after the AFM transition as we can see the change in MnPSe$_3$ bands in changing the magnetic states (Fig. S8-10). The band alignment, however, does not affect excitonic peak positions inside monolayer TMDs (which is not *interlayer* but *intralayer* exciton) as far as the bandgaps of TMDs themselves are not altered as mentioned above.

§10. Strain effect due to magnetostriction in MnPSe$_3$

Magnetostriction of the antiferromagnets below $T_N$ can compress the lattice of TMDs, which may potentially explain the observed upshift through a strain effect. The observed shift (~ 5 meV) requires roughly 0.1% compressive strain[14,15]. However, a previous study on bulk MnPSe$_3$ showed that the magnetic strain is, if any, one order of magnitude smaller (< 0.02%, below their measurement limit) across $T_N$ except the usual thermal expansion[1]. This suggests that the strain induced by magnetostriction on MnPSe$_3$ cannot explain our experimental results.

To confirm this, we performed the DFT calculations while changing the in-plane lattice constant (Fig. S11). As shown in Fig. S11a, the bandgap of MoS$_2$ is changed linearly to the lattice constant. Figure S11b shows the total energy comparison of the systems with different lattice constants for different $U$ in the GGA+$U$ calculations. The result indicates that the larger electron interaction leads to the larger lattice constant in the stable crystalline structure. Since $U$ stabilizes the Néel order, our results show that the additional strain from the Néel ordering is expected to be tensile, inducing the downshift opposite to the experimental data. Thus, the magnetic strain cannot rationalize our experimental results.

§11. Magnetic polaron effect at the magnetic van der Waals heterointerface

The exciton binding energy in monolayer TMDs could be modified due to the magnetic ordering in MnPSe$_3$. Magnetic-polaron picture is frequently used to describe the excitonic features in magnetic semiconductors[20]. This picture, however, also results in the opposite peak shift since the binding energy of excitons gets larger due to the magnetic polaron effect.

Meanwhile, two elementary excitations composing polaronic states can generally interact each other not only attractively (similarly to the bound magnetic polaron in magnetic semiconductors) but also repulsively[21]. If there is strong repulsion between excitonic (high energy) and magnonic (low energy) states in our system, it can explain the upshifts of the excitonic states. Because this mechanism can include exciton-magnon coupling in a broad sense, we don't mention it explicitly in the main text.

The effects on the mass of exciton in TMDs from antiferromagnetic MnPSe$_3$ is also not likely because the band edge of TMD, which determines the mass of exciton, is not so modulated directly by MnPSe$_3$ as discussed in §7 and §8.

§12. Interlayer exciton-magnon coupling at the heterointerfaces

In the main text, we attribute the observed shift to the effect of the interlayer exciton-magnon interaction. According to previous studies in a single material, the form of Hamiltonian inducing the exciton-magnon interaction ($\mathcal{H}_{\text{EMC}}$) under radiative electric field ($\vec{E}$) can be written as

$$\mathcal{H}_{\text{EMC}} = (\vec{\pi} \cdot \vec{E})a^\dagger c^\dagger + h.c. \qquad (2)$$

where $\vec{\pi}$ is coupling constant related to the exchange interaction between the ions, $a^\dagger$ is the exciton creation operator, and $c^\dagger$ is the magnon creation operator[22,23]. This original theory on exciton-magnon state in d-d transition is based on the electric dipole moment generated by the combination of orbital excitation (exciton) at one sublattice and spin excitation (magnon) at the other sublattice. The dipole moment makes the strong optical transition allowed and strong sideband peaks appear below the $T_\text{N}$ [22,23]. This theory is similar to the theory on two-magnon process (spin excitations at both sublattices) and exciton-phonon coupling.

In the interlayer EMC, naïvely thinking, we can regard $a^\dagger$ as the exciton creation operator at the TMD layer, $c^\dagger$ as the magnon creation operator at the MnPSe$_3$ layer, and $\vec{\pi}$ as interlayer coupling constant linked to the exchange coupling between TMDs and MnPSe$_3$. Note that the exchange couplings between TMDs and ferromagnetic surfaces were confirmed in the recent studies as mentioned in the main text[24,25].

However, it is challenging to construct an appropriate theory in more detailed. First, although the exciton-magnon states in d-d transitions (electrical-dipole forbidden/magnetic-dipole allowed) are quite well understood, those in charge transfer transition (electrical-dipole allowed, same as exciton in TMDs) are much less investigated even in bulk as far as we know. It would be quite important to establish the detailed theory about exciton-magnon states in electrical-dipole allowed transitions with originally large oscillator strength.

The original formula (2) is originally made for an exciton and a magnon at localized d-orbitals in one substance. In the heterointerface, however, the magnon and exciton exist in separated layers and are supposed to interact via interlayer exchange coupling. Moreover, the exciton in TMDs is Wannier excitons which is not localized at the atomic sites and distinct from Frenkel excitons in localized d-electron system. These situations can make it complicated and/or interesting to construct the microscopic theory of the interlayer exciton-magnon coupling.

As a result of (2), energy conservation leads to the simple relation in the optical transition:

$$E_\text{photon} = E_\text{exciton} \pm E_\text{magnon}. \qquad (3)$$

The observed upshift in the luminescence can be explained by the case of the upper sign that a photon is created while an exciton and a magnon are annihilated; however, the downshift (corresponding to the lower sign, a photon and a magnon are created and an exciton is annihilated) can be also observed in general. Although it is difficult to conclude the reason why we cannot detect the downshift apparently, the upshift could dominate under the sufficient population of thermally and/or optically excited magnons as reported in a bulk $MnF_2$[26].

It would be also fascinating to discuss the magnetic field dependence to unveil detailed features of the interlayer exciton-magnon coupling. In d-d transitions of $Mn^{2+}$-based bulk materials, it is known that exciton-magnon peaks show almost no Zeeman splitting due to the cancellation coming from the coincidence of optically and magnetically excited states inside $Mn^{2+}$ orbitals[27,28]. In our case, although there is no comprehensive theory, we can imagine that Zeeman shifts inside $Mn^{2+}$ orbitals can be observed adding to or subtracting from the usual Zeeman splitting in TMDs since no exact cancellation would occur in our interlayer system now. Because the Zeeman splitting in TMDs dependents on the valley index (valley Zeeman shift), the g-factor of such a magnon-coupled valley-exciton would be worth discussing from the theoretical and experimental viewpoints.

Moreover, it would be intriguing to measure such field dependence both in Faraday and Voigt geometries because it would give us the exact easy-axis of the sublattice magnetic moments just at the interface of $MnPSe_3$ (which can be different from the bulk). This method can be a tool for detecting the surface magnetic ordering of antiferromagnets at heterointerfaces.

a. Parallel 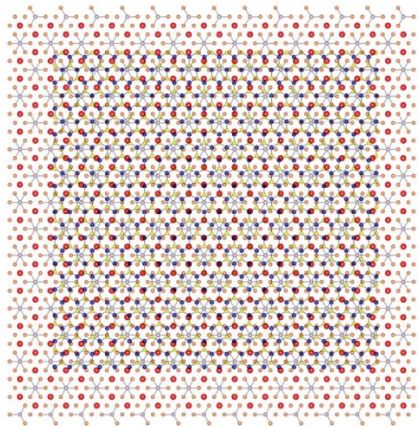

b. Perpendicular 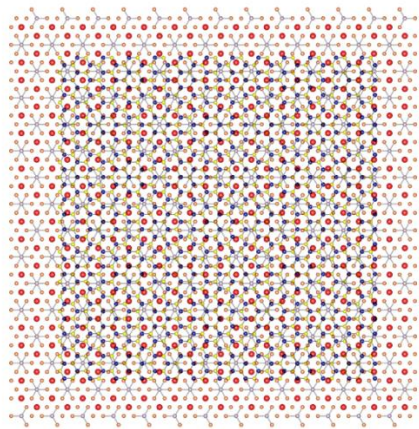

**Fig. S1. Stacking images.** Top views of MoSe$_2$/MnPSe$_3$ with different stacking angles: the zigzag-edge of MoSe$_2$ parallel to the zigzag-edge of honeycomb Mn$^{2+}$ (same as sample A1, A2, and A3) and the zigzag-edge of MoSe$_2$ perpendicular to the zigzag-edge of Mn$^{2+}$ (same as sample E1 and E2). Large and small squares are monolayer MnPSe$_3$ and MoSe$_2$ films, respectively. The color of each element is same as Fig. 1a except for the selenides in MoSe$_2$ (yellow here).

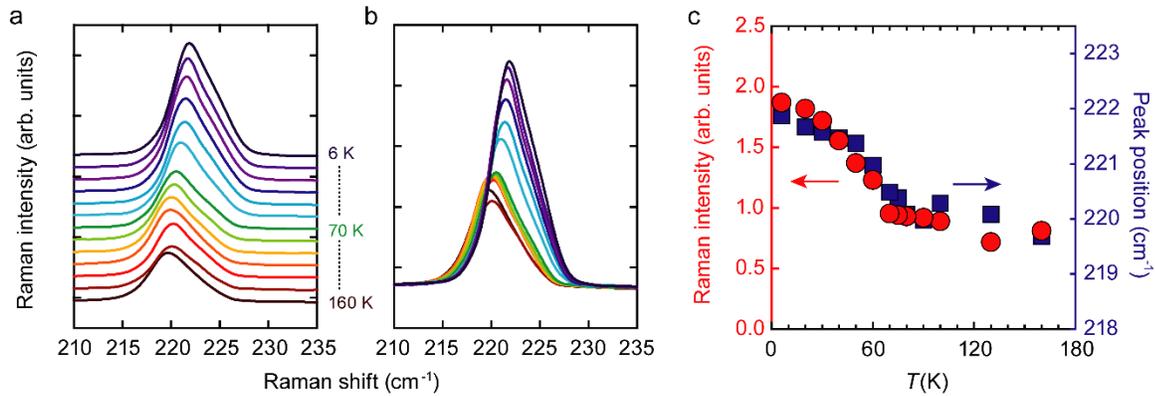

**Raman peak around 222 cm$^{-1}$ from MnPSe$_3$. a**, **b**, Raman spectra around 220 cm$^{-1}$ from MnPSe$_3$ at 6, 20, 30, 40, 50, 60, 70, 75, 80, 90, 100, 130, and 160 K. **c**, Temperature dependence of intensities and positions of the peaks in Fig S2b. Their behaviors change around $T_N$ (= 74 K) of the bulk MnPSe$_3$.

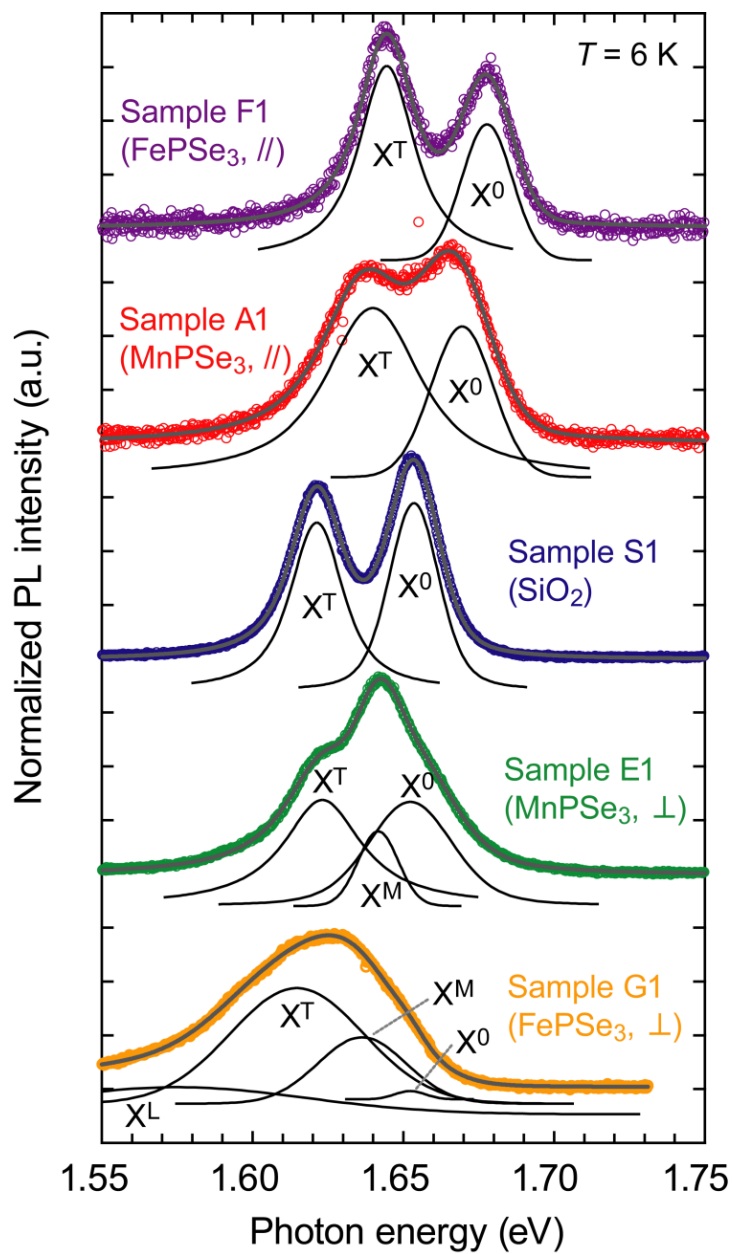

**Fig. S3. PL spectra of all MoSe$_2$/$M$PSe$_3$ interfaces.** PL spectra from monolayer MoSe$_2$ in sample F1 (purple), A1 (red), S1 (blue), E1 (green), and G1 (yellow) with the results of the multi-peak fittings by Voigt function. Grey and black lines show the fit traces and peaks for each spectrum, respectively.

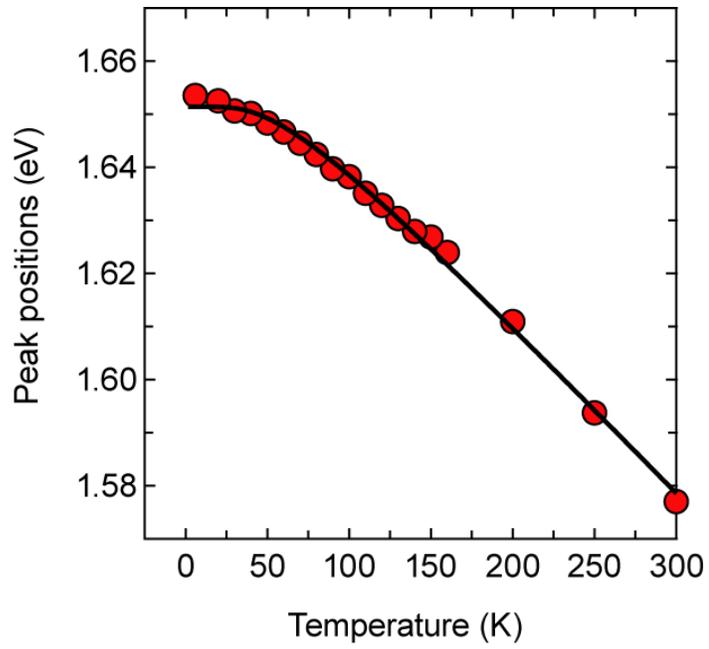

**Fig. S4. Temperature dependence of PL peaks in MoSe$_2$ on SiO$_2$.** The PL peaks of X0 of sample S1 are plotted as a function of temperature with the fitting by (1).

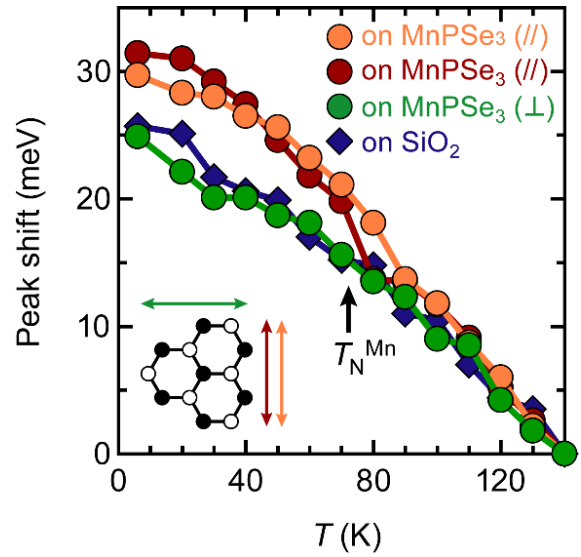

**Fig. S5. Temperature dependence in other MoSe$_2$/MnPSe$_3$ samples.** The PL peak shifts from different MoSe$_2$ samples from Fig. 4d. Orange and Brown circles correspond to the data from sample A2 and A3 of MoSe$_2$ on MnPSe$_3$ with the parallel configuration. Green circles and blue rhombuses come from sample E2 of MoSe$_2$ on MnPSe$_3$ with the perpendicular configuration and sample S2 on MoSe$_2$ on SiO$_2$, respectively.

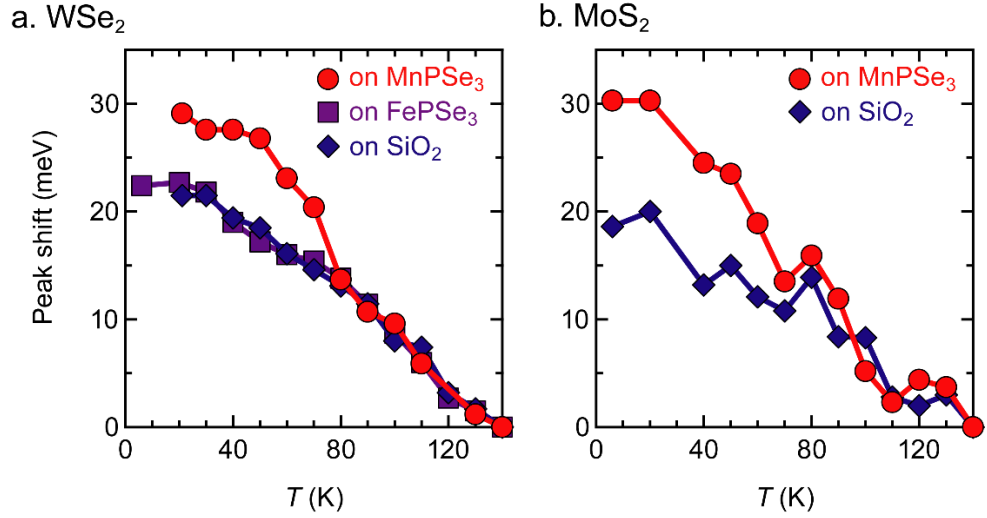

**Fig. S6. Peak shifts in the heterointerfaces using WSe$_2$ and MoS$_2$.** Summary of the shifts of PL peaks from **a** WSe$_2$ and **b** MoS$_2$. Red circles, purple squires and blue rhombuses from TMDs on MnPSe$_3$, FePSe$_3$ and SiO$_2$ respectively. Note that the PL peaks from MoS$_2$ are much broader than others[29], which make it difficult to deduce the peak positions precisely by fitting.

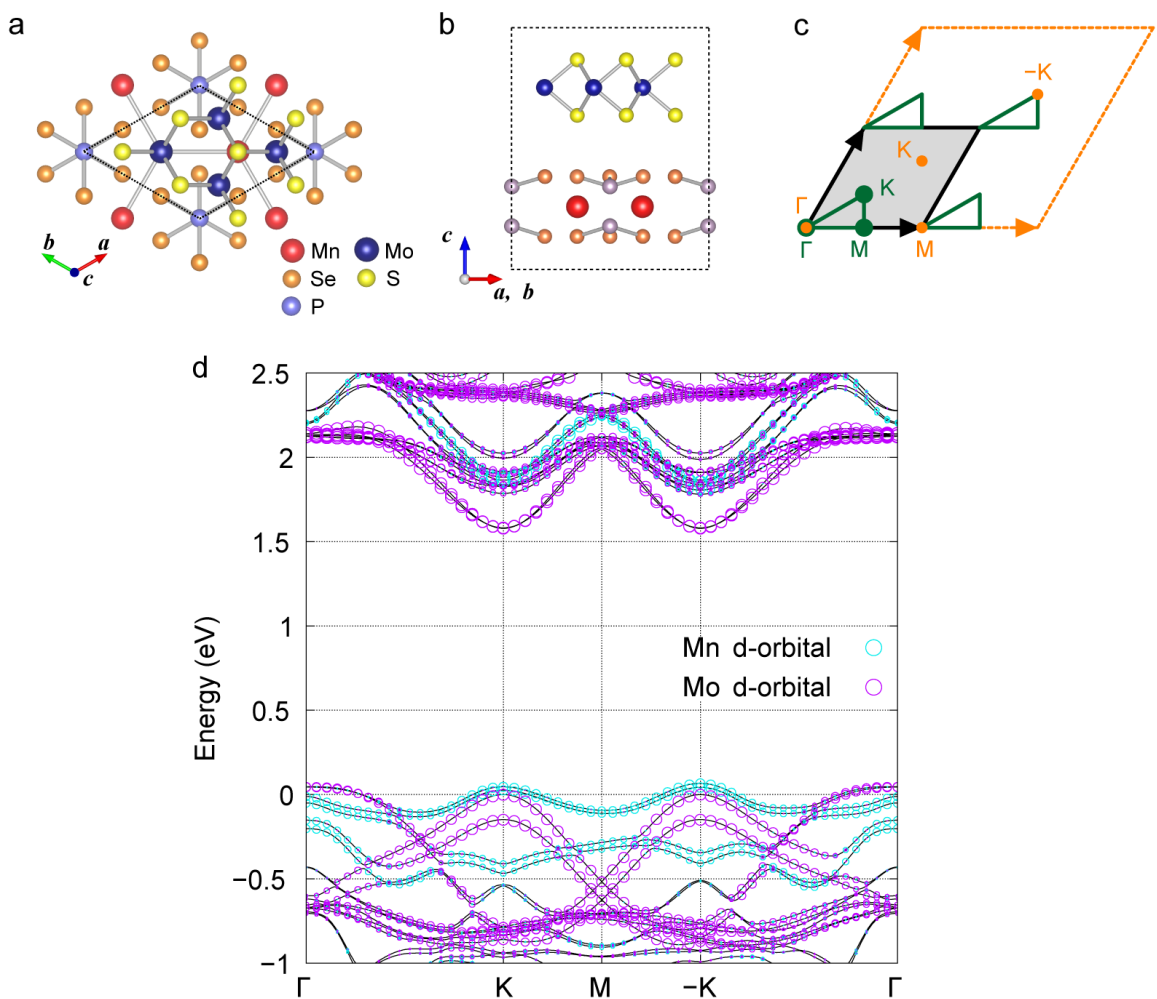

**Fig. S7. Electronic band structure of MoS$_2$/MnPSe$_3$. a, b,** Top and side views of the unit cell of the MoS$_2$/MnPSe$_3$ bilayer system. **c,** 1st Brillouin zone (BZ) and high-symmetry points for the MoS$_2$ itself (orange rectangle and letters) and MoS$_2$/MnPSe$_3$ superlattice (black rectangle and green letters). The green lines indicate the symmetric lines used for the plots of the band structures in the following. Note that the Γ, K, and M points are the high-symmetry points of the superlattice written in green letters in **c** hereafter. **d,** Electronic band structure of the bilayer system obtained by the GGA+$U$ calculations ($U$ = 5 eV). The radius of cyan and magenta circles indicates the weight of the *d* orbitals of Mn and Mo, respectively (the Mn-components are multiplied by four for clarity). We set the origin of the energy at the top of the valence band of $MoS_2$.

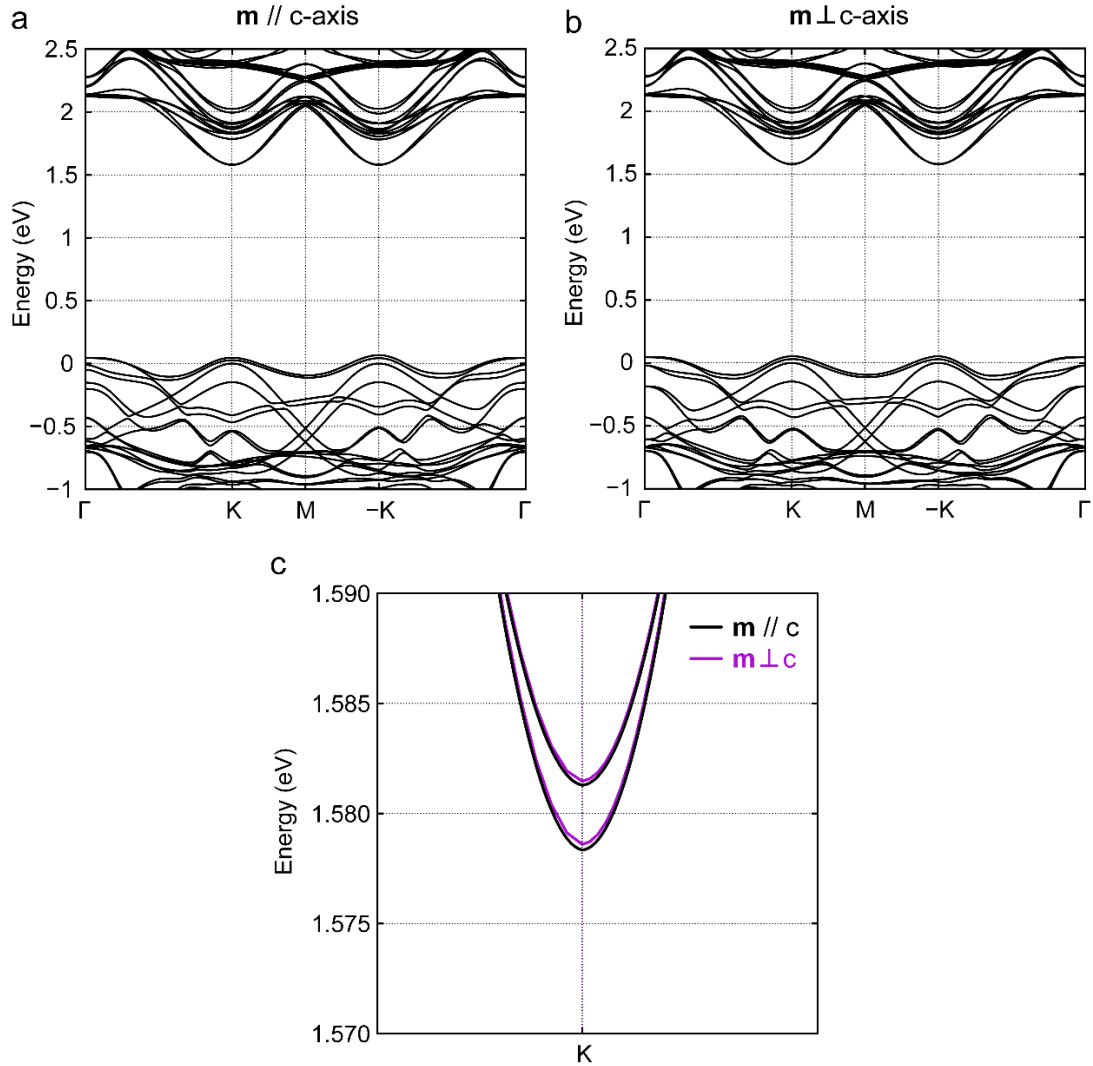

**Fig. S8. Magnetization direction dependence of electronic band structures. a, b,** Electronic band structures obtained by the GGA+$U$ calculations ($U$ = 5 eV) with Néel ordering of **a m**//c-axis and **b m**⊥c-axis. **c**, Enlarged figure of **a** and **b** around the conduction band bottom of MoS$_2$. The difference of the bandgaps (< 0.5 meV) is one order of magnitude smaller than the upshift of the PL spectra observed in experiments (~5 meV).

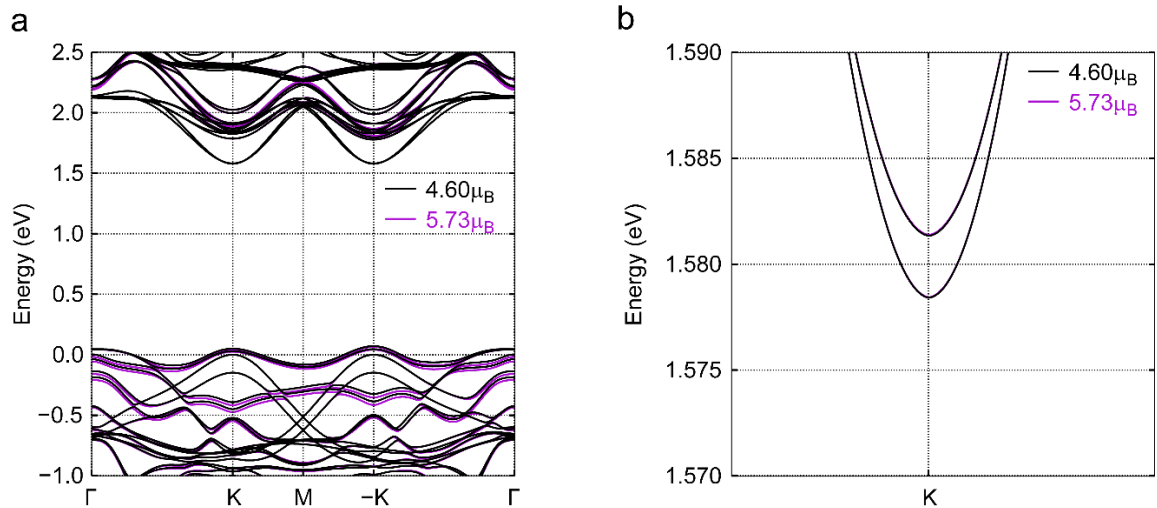

**Fig. S9. Magnetic moment dependence of electronic band structures. a,** Electronic band structures obtained by the GGA+$U$ calculations ($U$ = 5 eV) with the constraints on the magnetic moments of Mn. The black (purple) line shows the results with the magnetic moment $m$ = 4.60 (5.73) $\mu_B$ along the c axis on each Mn site. **b,** Enlarged figure of **a** around the conduction band bottom of $MoS_2$. The deduced bandgaps of $MoS_2$ are almost identical between the two cases.

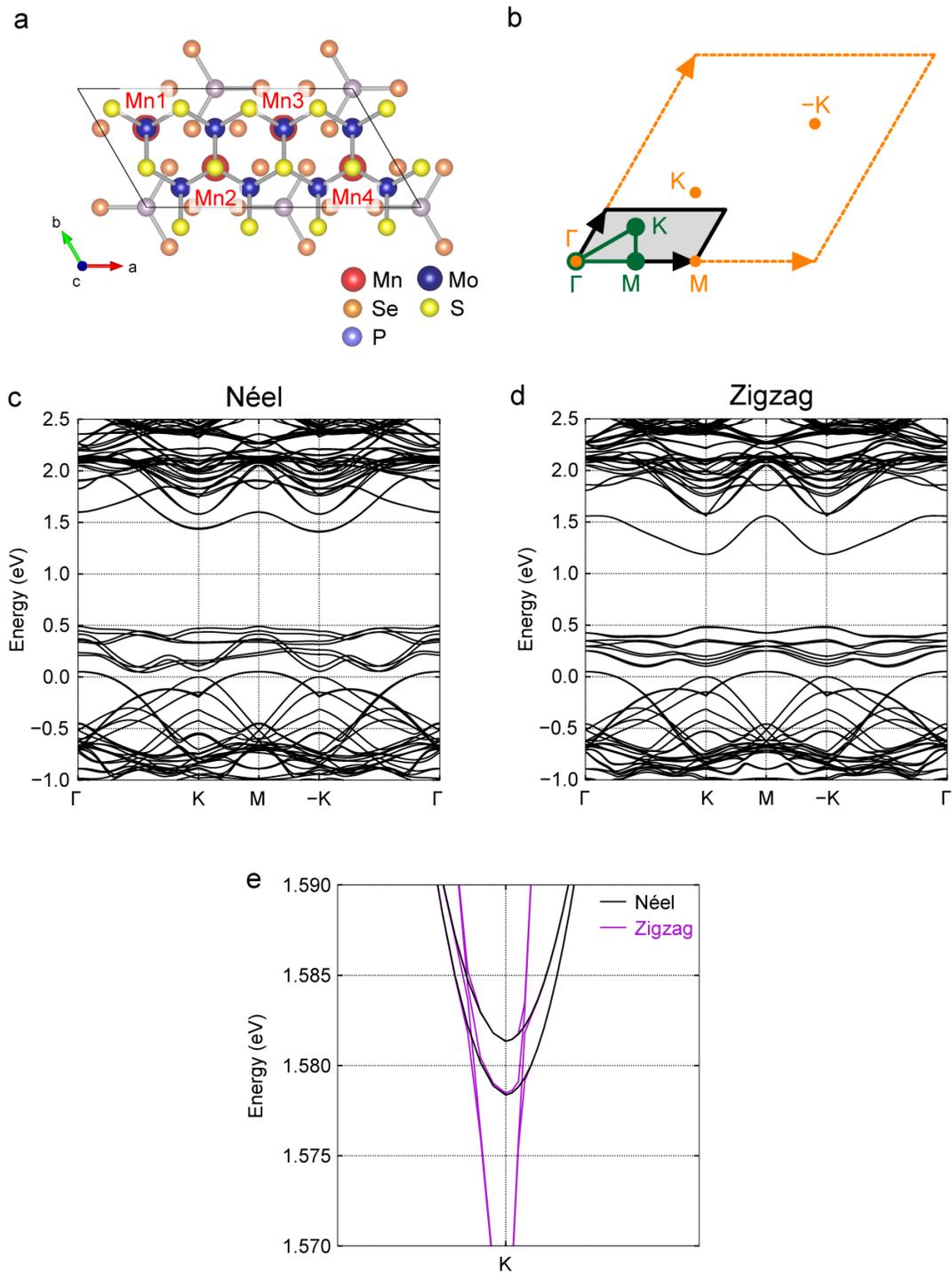

**Fig. S10. Antiferromagnetic structure dependence of electronic band structures. a,** The rectangular double supercell including four $Mn^{2+}$ sites (denoted by Mn1, 2, 3, and 4). **b,** 1st Brillouin zone (BZ) for $MoS_2$ (orange rectangle) and $MoS_2$/$MnPSe_3$ double supercell in **a** (black

rectangle). Here, we use the same Γ, K, and M points as in Fig. S7c for comparison. Note that the green lines, which are used for the plots of the band structures in **c** and **d**, are not the symmetry lines for the BZ for this double supercell. **c, d,** Electronic band structures obtained by the GGA calculations with **c** Néel order [Mn1 and 3 (Mn2 and 4) possess up (down) spin along the c axis] and **d** zigzag order [Mn1 and 2 (Mn3 and 4) possess up (down) spin along the c axis]. The kinks at the K points are due to the fact that the Γ-K-M line is not the symmetry line for the doubled supercell. **e,** Enlarged figure around the conduction band bottom of $MoS_2$. The band structures from $MoS_2$ are almost identical except for the additional bands from Mn *d* orbitals in the case of zigzag order.

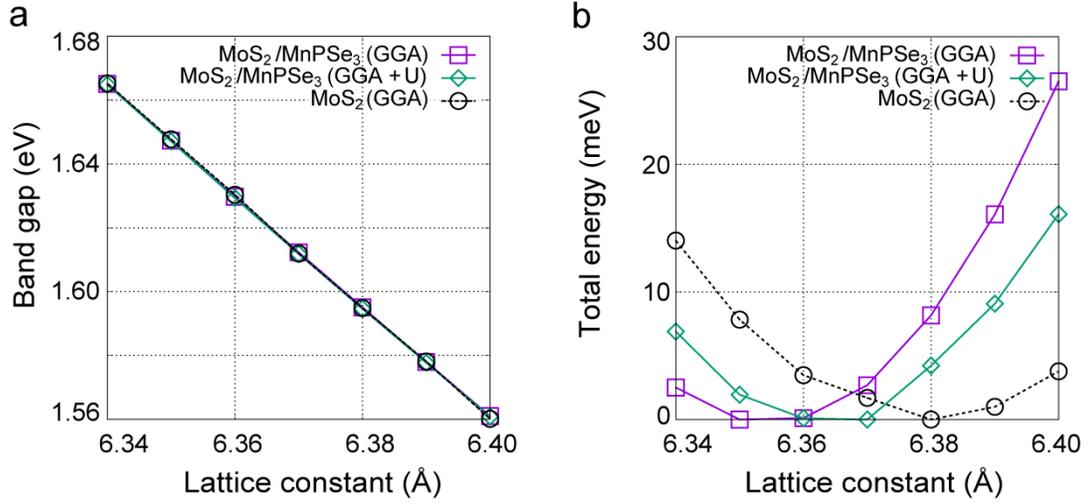

**Fig. S11. Lattice constant dependence of bandgap of TMD. a**, **b**, Bandgap of $MoS_2$ and total energy as functions of the lattice constant. The purple, green, and black lines indicate the results by the GGA calculations for the bilayer $MoS_2/MnPSe_3$, the GGA+$U$ calculations ($U = 2$ eV) for the bilayer $MoS_2/MnPSe_3$, and the GGA calculations for the monolayer $MoS_2$, respectively. In **b**, the total energy is measured from the lowest total energy for each case.